\documentstyle[preprint,aps,epsfig,axodraw]{revtex} 
\newcommand{\lsim}{\mathrel{\mathop{\kern 0pt \rlap 
  {\raise.2ex\hbox{$<$}}} 
  \lower.9ex\hbox{\kern-.190em $\sim$}}} 
\newcommand{\gsim}{\mathrel{\mathop{\kern 0pt \rlap 
  {\raise.2ex\hbox{$>$}}} 
  \lower.9ex\hbox{\kern-.190em $\sim$}}} 
\newcommand{\lm}     {\lambda} 
\newcommand{\Lm}     {\Lambda} 
\newcommand{\gm}     {\gamma} 
\newcommand{\beq}     {\begin{equation}} 
\newcommand{\eeq}     {\end{equation}} 
\newcommand{\bea}     {\begin{eqnarray}} 
\newcommand{\eea}     {\end{eqnarray}} 
\newcommand{\M}     {{\mathcal M}} 
\begin{document} 
\draft 
\preprint{ 
\vbox{\hbox{\bf (hep-ph/0009231)} 
      \hbox{YUMS 00-09, ~~KIAS P00060, ~~SNUTP\hspace*{.2em}00-024}} 
} 
\title{ 
Enhancement of the Higgs pair production at the LHC;\\ 
the MSSM and extra dimension effects 
} 
\author{ 
C.~ S.~ Kim$^{\:a,b}$,~~ Kang~ Young~ Lee$^{\:c}$~~ and~~ Jeonghyeon~ 
Song$^{\:c}$ } 
\vspace{1.5cm} 
\address{ 
$^a$Department of Physics and IPAP, Yonsei University, Seoul 120-749, Korea\\ 
$^b$Department of Physics, University of Wisconsin, Madison, WI 53706, USA\\ 
$^c$School of Physics, 
Korea Institute for Advanced Study, Seoul 130-012, Korea
} 
\maketitle 
\thispagestyle{empty} 
\setcounter{page}{1} 
 
\begin{abstract} 
\noindent 
The neutral Higgs pair production at the LHC 
is studied in the MSSM, the large extra dimensional (ADD) model and 
the Randall-Sundrum (RS) model, 
where the total cross section can be significantly 
enhanced  compared to that in the SM. 
The $p_{_T}$, invariant mass and rapidity 
distributions of each model have been shown 
to be distinctive: 
The ADD model raises the $p_T$ and invariant mass distributions 
at high scales of $p_T$ and invariant mass; 
in the RS model resonant peaks appear after the SM contribution dies away;
the SM and the MSSM distributions 
drop rapidly at those high scales; in the ADD and the RS models 
the rapidity distributions 
congregate more around the center. 
It is concluded that 
various distributions of the Higgs pair production 
at the LHC with restrictive kinematic cuts 
would provide one of the most robust signals 
for the extra dimensional effects.

\pacs{PACS numbers:  04.50.+h,12.60.-i,13.85.-t,13.90.+i } 
 
\end{abstract} 
 
\newpage 
 
\section{Introduction} 
 
The standard model (SM) has been very successful in explaining 
experimental signals, including recent results from the CERN  
LEP-II collider \cite{SM}. 
Nevertheless one of the most important ingredients of the SM, the 
Higgs mechanism, has not been experimentally probed yet; 
it is responsible for spontaneous electroweak 
symmetry breaking accompanying the mass generation of the $W^{\pm}$ and 
$Z^0$ gauge bosons as well as the SM fermions. 
Recently, ALEPH group of the LEP-II has reported the observation 
of an excess of 3$\sigma$ in the search of the SM Higgs boson, 
which corresponds to the Higgs mass about 114 GeV \cite{ALEPH}. 
As operations of the LEP-II have been completed, 
the decision, whether the observations are only the results of statistical 
fluctuations or the first signal of the Higgs boson production, 
is suspended until the Fermilab Tevatron II and/or 
the CERN LHC experiments running
\cite{Ellis}.  Thus it is naturally anticipated that primary 
efforts of future collider experiments are to be directed 
toward the search for Higgs bosons \cite{Higgs}. 
 
In particular at hadron colliders, 
the pair production of Higgs bosons holds a 
distinctive position in understanding the Higgs mechanism \cite{Zerwas}. 
First, 
it may provide the experimental reconstruction of the Higgs 
potential, as the triple self-coupling of Higgs particles is 
involved. 
The establishment of the Higgs' role in the electroweak 
symmetry breaking is crucially dependent on this measurement. 
Second, the 
signal-to-background ratio is significantly improved compared to that of 
a single Higgs boson production. 
The invariant mass scale of the $single$ Higgs production 
is fixed by the Higgs mass, of order $\sim 
100$ GeV. Thus their detection through heavy quark decay modes 
suffers from large QCD backgrounds. Besides, one viable decay mode 
$h \to \gamma \gamma$ has a very small branching ratio of order 
$10^{-3}$ \cite{hgg}. For the $pair$ production of the Higgs 
particles, the four $b$-jets in the final states are energetic, 
reducing the main background $h b \bar{b}$ with soft 
$b$-jets \cite{Belyaev}. Third, this is a rare process in the sense 
that the effects of physics beyond the SM can remarkably enhance 
the cross section with respect to that in the SM; the minimal 
supersymmetric standard model (MSSM) \cite{MSSM} provides 
some parameter space for the 
large enhancement of the total cross section, 
which should accommodate the large Yukawa coupling of the $b$ 
quarks, the resonant decay of $H\to h h $, and/or 
dominantly large contribution of the squark loops \cite{Belyaev}; 
extra dimensional models provide tree level diagrams mediated by 
the Kaluza-Klein (KK) gravitons, leading to large total 
cross sections. In fact, these new theoretical approaches have 
drawn extensive attention as  candidates for the solution of the 
gauge hierarchy problem, the existence and stability of the 
enormous hierarchy between the electroweak and Planck scales. 
Therefore, it is worth studying the production of a neutral Higgs 
pair with the effects of the MSSM and the extra dimensional models, 
and finding the characteristic distribution of each model. We 
shall restrict ourselves to the procedure at the LHC, which 
is scheduled to start operating in 2005.  
Despite the 
assurance of the detectibility of the lightest CP-even MSSM 
Higgs boson with $\sqrt{s} \gsim 250$\,GeV and 
$\int {\mathcal L} dt \gsim 10$\, fb$^{-1}$ \cite{Barger}, 
the LHC takes a practical 
advantage over the future $e^+e^-$ linear colliders  
which lack specific plans yet. 
 
The remainder of the paper is organized as follows. 
The next section details the 
neutral Higgs pair production at the 
LHC in the SM, the MSSM, 
the large extra dimensional (ADD) model \cite{ADD}, and the Randall-Sundrum 
(RS) model \cite{RS}. 
In Sec.~III, we discuss various distributions useful to 
discriminate effects of each model from others. 
The last section deals with a brief summary and conclusions. 
 
\section{Theoretical Discussions and Formulae} 
 
The production of a Higgs boson pair at a hadron collider proceeds 
through several modes; $W W$ fusion, bremsstrahlung of Higgs 
bosons off heavy quarks, and gluon-gluon fusion. At the LHC, the 
$gg$ fusion is expected to play a main role, since the gluon 
luminosity increases with beam energy. In this paper, we focus on 
the process $g g \to h h$, where the $h$ is the lightest Higgs 
boson in each theory.  
 
The invariant amplitude squared is generally 
written as, 
in terms of the helicity amplitude $\M_{\lm_1 \lm_2}$ 
for the initial gluon helicities $\lm_{1(2)}$ \cite{Belyaev}, 
\beq 
\overline{ \left| \M \right|^2} = 2 \cdot \frac{1}{4} \cdot 
\frac{1}{64} \cdot \frac{1}{2} \left[ \left| \M_{++} \right|^2 + 
\left| \M_{+-} \right|^2 + \left| \M_{-+} \right|^2 + \left| 
\M_{--} \right|^2 \right] , 
\eeq 
where the factor 2 refers  the 
color factor $[{\rm Tr} (T^a T^b)]^2=2$, the factor 1/4  the 
initial gluon helicity average, the factor 1/64  the gluon color 
average, and the final factor 1/2  the symmetry factor for the 
two identical Higgs boson production. For the processes with CP 
conservation,  the helicity amplitudes 
are related as $\M_{++} = \M _{--}$ and $\M_{+-} = \M_{-+}$. 
 
For the numerical 
analysis, we use the leading order MRST parton distribution 
functions (PDF) for a gluon in a proton \cite{MRST}. The QCD 
factorization and renormalization scales $Q$ are set to be the 
$h h$ invariant mass,  {\it i.e.}, $\sqrt{\hat{s}}$. 
The $Q^2$-dependence is expected to be small on the distribution shapes,
which are of our main interest.
The center-of-momentum (c.m.) energy at $pp$ collisions 
is $\sqrt{s}=14$ TeV. 
And we have employed the 
kinematic cuts of $p_{_T} \geq 25$ GeV and $|\eta| \leq 2.5$ throughout 
the paper. 
 
\subsection{In the SM} 
 
In the SM, there are two types of Feynman diagrams as depicted in 
Fig.~1. One is the triangle diagram where a virtual Higgs 
boson, produced from $gg$ fusion through heavy quark triangles, 
decays into a pair of Higgs bosons. The other is the box 
diagram where the Higgs pair is produced through heavy quark 
boxes. It is to be noted that the triangle diagram incorporates 
the triple self-coupling of Higgs bosons. For analytic 
expressions of all the one-loop helicity amplitudes of the process $g g \to 
h h$, we refer the reader to Refs. \cite{Belyaev,Zerwas}. Fig.~2 
shows the total cross section of the SM Higgs pair production at the LHC,
as a function of the Higgs mass $m_h$. 
At $m_h \simeq 100$ GeV, the $\sigma_{\rm tot}$ 
is of order 60 fb, which decreases rapidly with increasing 
$m_h$. 
 
\subsection{In the MSSM} 
 
The existence of a fundamental scalar particle in the SM causes 
the well-known gauge hierarchy problem. 
Traditional approaches are 
to introduce new symmetries, motivated by the chiral symmetry for 
light fermion masses and the gauge symmetries for gauge boson 
masses. Supersymmetry is one of the most popular candidates for 
this new symmetry. 
 
The MSSM Higgs pair production undergoes distinctive contributions 
coming from the new Higgs and squark sectors, which provide the 
possibilities to greatly enhance the total cross section of the 
process. First, two doublets and thus two different vacuum 
expectation values of the Higgs fields allow the Yukawa coupling 
of the $b$ quark to be compatible with that of the top quark, 
which corresponds to the large $\tan\beta$ case. 
Much smaller mass of the $b$ quark may makes its loop contribution even larger 
than the top quark contribution \cite{Zerwas}. Second, two Higgs 
doublets imply the presence of a heavy CP-even neutral Higgs 
boson $H$, which can  decay into two light Higgs bosons 
if kinematically allowed. This resonant contribution $gg \to H \to h h$ 
is shown to enhance the total 
cross section with respect to the SM case by about an order of 
magnitude. 
Third, the MSSM permits 
a parameter space where the $\tilde{b}$ or $\tilde{t}$ loop 
contributions can exceed the SM quark loop contributions by more 
than two orders of magnitude \cite{Belyaev}. For the maximization 
of squark loop contributions which occurs through the $\tilde{b}$ 
loops, this parameter space should allow large value of 
$\tan\beta$, considerably light $\tilde{b}_1$ mass, and 
large mass of $A$ and/or $|\mu|$. Since the third 
enhancement possibility has rather restrictive parameter space 
(for example, the squark loop contributions are practically 
negligible unless $m_{\tilde{b}_1} \lsim 120$ GeV), we consider 
only the quark loop contributions, as in Ref. \cite{Zerwas}. The 
corresponding Feynman diagrams are the same as the SM case, except 
for the different coupling strengths and the presence of the 
neutral heavy Higgs boson $H$ (see Fig.~1). 
 
The MSSM total cross section 
as a function of the Higgs mass shows similar behavior to that of the SM, 
except for the overall enhancement, 
as shown in Ref. \cite{Zerwas}. 
As anticipated, the total cross sections in the large and small 
$\tan\beta$ cases are much increased compared to 
the SM case. 
In particular, the large $\tan\beta$ value 
with $m_h \simeq 100$ GeV leads to 
an order of magnitude enhancement of the cross section. 
 
\subsection{In the ADD model} 
 
The gauge hierarchy problem 
has been approached 
without resort to any new symmetry 
by Arkani-Hamed, Dimopoulos, and Dvali (ADD) \cite{ADD}. 
One prerequisite of the hierarchy problem itself 
is removed; 
the Planck mass is not fundamental; 
the nature allows 
only one fundamental mass scale $M_{\rm S}$ 
which is at the electroweak scale. 
By introducing the $N \geq 2$ extra dimesional compact space, 
the observed huge Planck mass 
is attributed to 
the large volume of the extra space, 
since $M_{\rm Pl}^2 \simeq M_{S}^{N+2} R^N$ 
where the $R$ is the size of the extra dimension. 
The prohibition of the SM particles' 
escaping into the extra space 
can be achieved by describing 
the matter fields as open strings of which 
the end-points are fixed 
to our four-dimensional world. 
 
Of great interest and significance is 
that this idea is testable at colliders. 
The KK reduction from the whole $(4+N)$-dimensions 
to our four-dimensional world 
yields towers of massive KK-states 
in the four-dimensional effective theory. 
Even though the coupling of a KK-graviton 
to the ordinary matter fields 
is extremely suppressed by the Planck scale, 
tiny mass-splitting $\Delta m_{\rm KK} \sim 1/R$ 
(which is about $10^{-3}$ eV for the $N=2$ case) 
induces summation over all KK-states, 
which compensates for the Planck scale suppression. 
In addition to single graviton emission processes 
as missing energy events \cite{Peskin}, 
the indirect effects of the massive graviton exchange 
on various collider experiments \cite{ADD_phen}, 
and the possible Lorentz and CPT invariance violations through 
the change of the metric on the brane \cite{cskim} 
have been extensively 
studied.

For the Higgs pair production through the gluon-gluon fusion, 
there exists a tree level Feynman diagram mediated 
by spin-2 KK-gravitons (see Fig. 3). 
Based on the effective four-dimensional Lagrangian \cite{han,Wells}, 
the helicity amplitudes are obtained as
\bea 
\M _{++} 
&=& 
\M _{--} =0 ~,
 \\ \nonumber 
\M_{+-} 
&=& 
\M_{-+} 
= 
\frac{ 16 \lm}{M_S^4} 
(m_h^4 - \hat{u} \hat{t} ) 
\,, 
\eea 
which are to be added to those in the SM. 
The interference effects between the ADD and the SM 
are proportional to $1/M_S^4$, 
which are, at the energy scale below $M_S$, 
dominant over pure ADD effects. 
 
There might be a concern with our ignorance of the parton mode 
$ q \bar{q} \to h h $ mediated by the KK-gravitons. 
The concern appears reasonable: 
The characteristic parton energy scale $\sqrt{\hat{s}}$
of the process in extra dimensional 
models is of the order $M_{S} \sim$ TeV, 
unlike a few hundred GeV scale in the SM and the MSSM; 
the dominant momentum fraction $x$ may not be so small as in the 
SM and the MSSM cases and the magnitude of 
the parton distribution functions of, 
in particular, the valence quarks becomes substantial. 
In the following, we have taken into account of the parton mode 
$ q \bar{q} \to h h $, 
which has the scattering amplitude squared as 
\begin{equation} 
\label{Amp_qq} 
\overline{ | {\mathcal M}|^2 } 
(q \bar{q} \to h h) 
= \frac{1}{9 M_{ S}^8} (\hat{t}-\hat{u})^2 (\hat{t}\,\hat{u} - m_h^4) 
\,. 
\end{equation} 
According to our numerical analysis, 
the contribution of $ q \bar{q} \to h h $ mode 
turns out to be at most a few percent to the total cross sections, because 
the amplitude squared itself is smaller than that of 
$g g \to h h $ mode (by a factor of about 1/36) 
while, at TeV $\sqrt{\hat{s}}$, the PDF of a valence quark and a sea quark 
is of the same order with that of two gluons. 
 
Note that 
in the parton c.m. frame, 
the nonzero helicity amplitude can be written as 
\begin{equation} 
\label{amp_ADD} 
\left. \M_{+-} 
\right|_{\rm g g - c.m.} =- \frac{16\lm}{M_S^4}\, p_T^2\, \hat{s} ~, 
\end{equation} 
where the $p_{_T}$ is the transverse momentum of an outgoing 
Higgs particle. 
Unitarity is apparently violated at high energies, 
which is expected from the use of an effective Lagrangian. 
It is reasonable that we only consider 
the region where our perturbative calculations 
are reliable, 
which can be achieved by excluding the region with 
high  invariant mass. 
In Ref.~\cite{Han},  the partial wave amplitudes 
of the elastic process $\gm \gm \to 
\gm \gm $ in the ADD model 
have been examined, 
yielding the bound on the ratio $M_S/\sqrt{\hat{s}} \,$, and
a valid region is conservatively found to be $\sqrt{\hat{s}} \leq 0.9 M_S$. 
In the following analyses,  
we impose an additional kinematic bound such as 
$$ 
M_{hh} < 0.9~ M_S ~.
$$

In Fig.~2, we present the total cross 
section as a function of $m_h$ with $M_{ S}=2.5$ TeV. 
As expected from the presence of a tree level diagram in the ADD model, 
the total cross section is substantially increased with respect to the SM
case.  For a pair production of Higgs bosons with mass $115$~GeV, 
we have obtained 
\beq 
\frac{ 
\sigma_{\rm SM+ADD} (m_h=115 ~{\rm GeV}, M_S=2.5~{\rm TeV}) 
}{ 
\sigma_{\rm SM} (m_h=115 ~{\rm GeV}) 
} 
\approx 
1.72 
. 
\eeq 
We note that the ADD effects lead to a gentle drop-off 
of the $\sigma_{\rm tot}$ with respect to the Higgs boson mass, 
contrary to the rapid decrease in the SM. 
Thus, if the Higgs mass is large, 
the ADD model could still produce substantially large number of 
Higgs pairs at the LHC unlike in the SM case. 
 
\subsection{In the RS model} 
 
More recently, Randall and Sundrum (RS) 
have proposed another extra dimensional scenario 
where, without the $large$ volume of the extra dimensions, 
the hierarchy problem is solved by 
a geometrical exponential factor, called a warp factor \cite{RS}. 
The spacetime in this model has a single $S^1/Z_2$ 
orbifold extra dimension with the metric 
\begin{equation} 
d s^2= e^{-2 k r_c |\phi| } \eta_{\mu\nu} d x^\mu d x^\nu 
+ r_c^2 d \phi^2, 
\end{equation} 
where the $\phi$ is confined to $0 \leq |\phi| \leq \pi$. 
The $r_c$ is the compactification radius 
which is to be stabilized by an appropriate 
mechanism \cite{gw}. 
Two orbifold fixed points accommodate two three-branes, 
the hidden brane at $\phi=0$ and 
our visible brane at $|\phi|=\pi$ or {\it vice versa}. 
The allocation of our brane at $|\phi|=\pi$ 
renders a fundamental scale $m_0$ to appear as 
the four-dimensional physical mass $m=e^{-k r_c \pi} m_0$, 
which answers the hierarchy problem. 
And the effective Planck mass is 
$$ 
M_{\rm Pl}^2 
=({M^3}/{k}) (1- e^{-2k r_c \pi} ) ~,
$$ 
where the $M$ is the five-dimensional Planck scale. 
Note that all of the $M_{\rm Pl}$, $k$, and $M$ are 
of the Planck scale. 
 
The compactification of the fifth dimension 
leads to the following interaction Lagrangian  
in the four-dimensional effective theory 
\cite{RS_G}, 
\begin{equation} 
{\mathcal L} = -\frac{1}{M_{\rm Pl}} T^{\mu\nu}(x) h^{(0)}_{\mu\nu}(x) 
-\frac{1}{\Lambda_\pi}  T^{\mu\nu}(x) \sum_{n=1}^\infty 
h^{(n)}_{\mu\nu}(x) 
\,, 
\end{equation} 
where $ \Lambda_\pi \equiv  e^{-k r_c \pi} M_{\rm Pl}$. 
Unlike almost continuous KK-graviton spectrum 
in the ADD model, 
we have one zero mode of the KK-gravitons 
with the coupling suppressed by the Planck scale, 
and the massive KK-graviton modes 
with the electroweak scale coupling 
$\Lambda_\pi$. 
The masses of the KK-gravitons are also at electroweak scale, 
given by \cite{KKmass}, 
\begin{equation} 
m_n=k x_n e^{-k r_c \pi}=\frac{k}{M_{\rm Pl}} {\Lambda_\pi} x_n 
\,, 
\end{equation} 
where the $x_n$'s are the $n$-th roots 
of the Bessel function of order one. 
 
The scattering amplitudes of 
the KK-mediated diagrams 
in the narrow width approximation 
can be derived from the ADD ones 
with the following replacement in Eq. (2) \cite{RS_G}: 
\begin{equation} 
\frac{\lm}{M_S^4} \longrightarrow 
-\frac{1}{8 \Lm_\pi^2} 
\sum_{n=1}^\infty 
\frac{1}{\hat{s}-m_n^2+i m_n \Gamma_n}, 
\end{equation} 
where 
the total decay width of the $n$-th KK-graviton 
is 
$ \Gamma_n = \rho m_n x_n^2 (k/M_{\rm Pl})^2$, and 
the $\rho$, fixed to be one, is a model-dependent parameter 
\cite{RS_G}. 
 
The observables based on the four-dimensional 
effective theory are determined by 
two parameters, 
$({\Lambda_\pi}, {k}/{M_{\rm Pl}})$. 
The value of ${k}/{M_{\rm Pl}}$ may be 
theoretically constrained 
to be less than about 0.1 \cite{kM}: 
The magnitude of the five-dimensional curvature, 
$R_5=-20\, k^2$, 
is required to be smaller than $M^2~(\simeq M_{\rm Pl}^2)$,
so that the classical RS solution derived from the 
leading order term in the curvature remains reliable. 
The ${\Lambda_\pi}$ is expected to be below 10 TeV 
in order to explain the hierarchy problem. 
Unlike the $M_S$ in the ADD case, 
the  ${\Lambda_\pi}$ does not play the role of 
a cut-off, relieving the concern 
of the unitarity violation. 
According to the phenomenological studies of 
the cross section of $e^+ e^- \to \mu^+ \mu^-$ 
in the RS model, 
only the case with large value of $k/M_{\rm Pl}$  hints 
the unitarity violation; 
even for $k/M_{\rm Pl}\sim 1$, 
unitarity violation can occur at c.m. energy of several TeV \cite{RS_G}; 
the current LEP-II experiments 
and the Tevatron run-I 
have provided a lower bound of 
${\Lambda_\pi}$ to be about 1.5 TeV 
in the case of $k/{M_{\rm Pl}}=0.1$.

In Fig.~2, we plot the total cross sections with respect to the Higgs mass 
within the RS model. 
We set $ {\Lambda_\pi}=3$ TeV and $  k/{M_{\rm Pl}}=0.1$. 
For the equity in comparing with the ADD case, 
we have employed the upper bound of $M_{h h} < 0.9 \, \Lm_\pi$. 
Though smaller than in the ADD case, 
the total cross section in the RS case is larger than  
that in the SM: 
\beq 
\frac{ 
\sigma_{\rm SM+RS} (m_h=115 ~{\rm GeV}, M_S=3~{\rm TeV}) 
}{ 
\sigma_{\rm SM} (m_h=115 ~{\rm GeV}) 
} 
\approx 
1.51 
. 
\eeq 
The rate of the drop-off of  $\sigma_{tot}$ against $m_h$ 
is similar to the SM case. 
 
In order to demonstrate the dependence of 
$ k/{M_{\rm Pl}}$ and $\Lm_\pi$, 
Fig.~4  shows the total cross section  within the RS model
as a function of $\Lambda_\pi$, 
considering three values of the ratio 
$ k/{M_{\rm Pl}}=0.01$, $0.1$ and $0.3$. 
The Higgs mass is set to be 100 GeV and the upper bound in $M_{h h}$ 
is not applied. 
As the $\Lambda_\pi$ increases, 
the $\sigma_{\rm tot}$ drops rapidly. 
And it can be seen that smaller value 
of the ratio $ k/{M_{\rm Pl}}$ produces 
larger cross section.  
This is due to that the amplitude squared in the narrow width approximation 
is inversely proportional to 
$( k/{M_{\rm Pl}} )^4$ at each resonance, which yields 
dominant contribution. 
 
\section{Numerical Discussions and Distributions} 
 
In the previous section, 
we have shown the possibilities that the 
Higgs pair production can be greatly 
enhanced at the LHC. 
In such a circumstance, 
it is worthwhile to search for appropriate 
distributions which enable us to distinguish 
the contributions of one model from others. 
In the numerical analysis of the distributions, we have 
employed the following parameters: 
The Higgs mass is set to be 100 GeV; 
in the MSSM, $\tan\beta=30$, 
$\mu=-~640$ GeV, $M_{\tilde{t}}=M_{\tilde{b}}=1000$ GeV, 
and $A_t=A_b=-~410$ GeV; 
the $M_{ S}$ in the ADD model is 2.5 TeV; 
in the RS model, $\Lambda_\pi=3$ TeV and $k/M_{\rm Pl}=0.1$.

In Fig.~5, we present the $p_{T}$-distributions 
in the SM and the ADD cases. 
While the SM $p_{_T}$-distribution peaks around 150 GeV and 
drops rapidly with increasing $p_T$, 
the ADD effects slowly draw the distribution up at high $p_T$ region. 
Note that the absence of the differential cross section 
at $p_T \gsim 1$ TeV is due the employment of the upper 
bound in the $M_{h h}$. 
Fig.~6 shows the  $p_{T}$-distributions 
in the SM and the RS cases. 
The presence of the Kaluza-Klein gravitons in the RS model 
leaves a clear shape of resonance. 
We caution readers that the $p_T$-axis is plotted by log-scale: 
Even if the $d \sigma^{\rm SM+ RS}/d p_T$ at $p_T > 300$ GeV, 
which practically vanishes in the SM, 
is smaller than that at $p_T < 300$ GeV by an order of magnitude, 
the extensive contribution at high $p_T$ region renders the total 
cross section substantially enhanced. 
In Fig.~7, we demonstrate the MSSM $p_T$-distribution 
for the large $\tan\beta$ case. 
It drops rapidly with increasing $p_T$ as in the SM case, 
while it peaks around 25 GeV, lower than the SM case. 
The magnitude of the differential cross section is 
shown about twenty times larger than that in the SM case. 
 
Fig.~8 illustrates the invariant mass distributions of the Higgs pair
in the SM and the ADD cases. 
The SM case, where the top quark loop contributions are dominant, 
peaks around the threshold $\sqrt{\hat{s}} \simeq 2 m_t$. 
The ADD effects gently raise the $M_{h h}$-distribution 
at high $M_{h h}$ region. 
It is expected that the blind application of the effective 
Lagrangian in Eq.~(\ref{amp_ADD}) without any upper bound 
in $M_{h h}$ would yield continual increase in the 
$M_{h h}$-distributions, 
which would show an apparent violation of unitarity. 
We display the  $M_{h h}$-distribution of the RS case 
in Fig.~9. 
The high resonance peak in addition to the SM distribution  
implies the first KK-state of 
gravitons with $m_1 \simeq 750$ GeV. 
While at the $e^+ e^- \to \mu^+ \mu^-$ process 
the KK-gravitons appear as almost regularly spaced peaks\cite{RS_G}, 
the hadronic convolution of the parton level processes 
obscures successive and separated peaks as 
the classical KK-signature. 
Fig.~10 displays the $M_{h h}$-distribution 
of the MSSM case with large $\tan 
\beta$, which undergoes 
dominant contributions from $b$ quarks. 
Thus a peak appears just above the kinematic threshold, 
$\sqrt{\hat{s}} \simeq 2 m_h$. 
 
Finally, we illustrate the rapidity distributions in Fig.~11. 
It can be seen that the extra dimensional models 
produce a Higgs pair somewhat more centrally in rapidity 
than the SM and the MSSM do. 
More restrictive cut on $\eta$, such as $\eta \leq 1.0$, 
would eliminate a substantial portion of the SM and the MSSM 
contributions.

\section{Conclusions} 
 
The pair production of neutral Higgs bosons 
from the gluon-gluon fusion at the LHC 
has been studied in the SM, the MSSM, the large extra dimensional 
model and the Randall-Sundrum model. 
We have shown that both the supersymmetric and extra-dimensional 
models can substantially enhance 
the total cross section of the Higgs pair production. 
In the MSSM case, 
the large $\tan\beta$ value makes possible the 
$b$-quark contribution dominant over the top quark contribution 
and the resonant decay of a heavy Higgs particle into two 
light Higgs particle. 
The extra dimensional models allow tree level 
Feynman diagrams mediated by the Kaluza-Klein gravitons, 
which significantly increase the total cross section. 
Since the ADD model has been shown to undergo 
the violation of partial wave unitarity 
at high energies, 
we have employed an upper bound in the invariant mass 
of two Higgs bosons, 
which is obtained from the analysis of the $J$-partial wave amplitudes 
of the elastic process $\gm \gm \to \gm \gm$. 
In addition, it was shown that 
the total cross section in the ADD case 
with the upper bound in $M_{h h}$ 
decreases gently with increasing Higgs mass, 
whereas those in the SM, the MSSM and the RS cases decrease rapidly. 
 
If Higgs pairs are produced at hadron colliders 
much more than the SM prediction, 
the three non-standard models considered here
are good candidates for the explanation. 
We have demonstrated the $p_{_T}$, invariant mass and rapidity 
distributions of each case. 
The distribution shapes are shown to be different for each model, 
providing valuable criteria to distinguish the 
contribution of one model from others. 
The $p_{_T}$-distribution in 
the SM peaks and drops rapidly; 
the MSSM one for the large $\tan\beta$ case 
peaks just above the threshold of $p_T$ 
and also drops rapidly; 
the ADD effects induce a slow raising after the SM peak; 
the RS effects can be discovered by 
the presence of a resonance peak. 
The invariant mass distributions 
are similar to the $p_{_T}$-distributions: 
The SM and the MSSM have peaks around a few hundred GeV; 
the ADD effects gently raise the $M_{h h}$-distribution 
at high energies; 
the RS contribution yields a series 
of resonant peaks. 
Therefore, restrictive cuts on the $p_T$ and $M_{h h}$ 
would eliminate the main contributions of the SM and the MSSM cases, 
which provides one of the most straightforward methods 
to signal the existence of low scale quantum gravity effects. 
Finally, the rapidity distributions 
in the ADD and the RS models 
show significantly narrow peaks around $\eta=0$,
which implies the large contributions at high $p_{_T}$ region.

\acknowledgments 
 
\noindent 
We thank G. Cvetic and P. Zerwas for careful reading of the manuscript and 
valuable comments. 
The work of C.S.K. was supported 
in part by  BK21 Program, SRC Program and Grant No. 2000-1-11100-003-1
of the KOSEF, and in part by the KRF Grants, Project No. 2000-015-DP0077. 
The research of J.S. was supported by the BK21 Program 
for the Seoul National University. 
 
\def\IJMP #1 #2 #3 {Int. J. Mod. Phys. A {\bf#1},\ #2 (#3)} 
\def\MPL #1 #2 #3 {Mod. Phys. Lett. A {\bf#1},\ #2 (#3)} 
\def\NPB #1 #2 #3 {Nucl. Phys. {\bf#1},\ #2 (#3)} 
\def\PLBold #1 #2 #3 {Phys. Lett. {\bf#1},\ #2 (#3)} 
\def\PLB #1 #2 #3 {Phys. Lett. B {\bf#1},\ #2 (#3)} 
\def\PR #1 #2 #3 {Phys. Rep. {\bf#1},\ #2 (#3)} 
\def\PRD #1 #2 #3 {Phys. Rev. D {\bf#1},\ #2 (#3)} 
\def\PRL #1 #2 #3 {Phys. Rev. Lett. {\bf#1},\ #2 (#3)} 
\def\PTT #1 #2 #3 {Prog. Theor. Phys. {\bf#1},\ #2 (#3)} 
\def\RMP #1 #2 #3 {Rev. Mod. Phys. {\bf#1},\ #2 (#3)} 
\def\ZPC #1 #2 #3 {Z. Phys. C {\bf#1},\ #2 (#3)} 
\def\EPJ #1 #2 #3 {Eur. Phys. J. C {\bf#1},\ #2 (#3)}

\smallskip 
\smallskip 
\smallskip 
 
\begin{center} 
\begin{figure}[htb] 
\hbox to\textwidth{\hss\epsfig{file=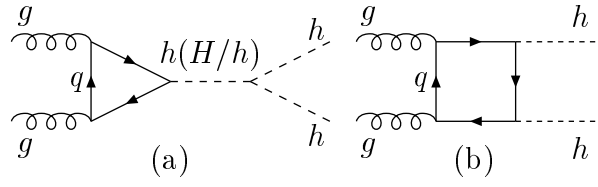}\hss} 
\caption{ The Feynman diagrams of the $gg \to h h$ process 
                in the SM (MSSM). 
} 
\label{fig1} 
\end{figure} 
\end{center} 
 
\newpage 
 
\begin{center} 
\begin{figure}[htb] 
\hbox to\textwidth{\hss\epsfig{file=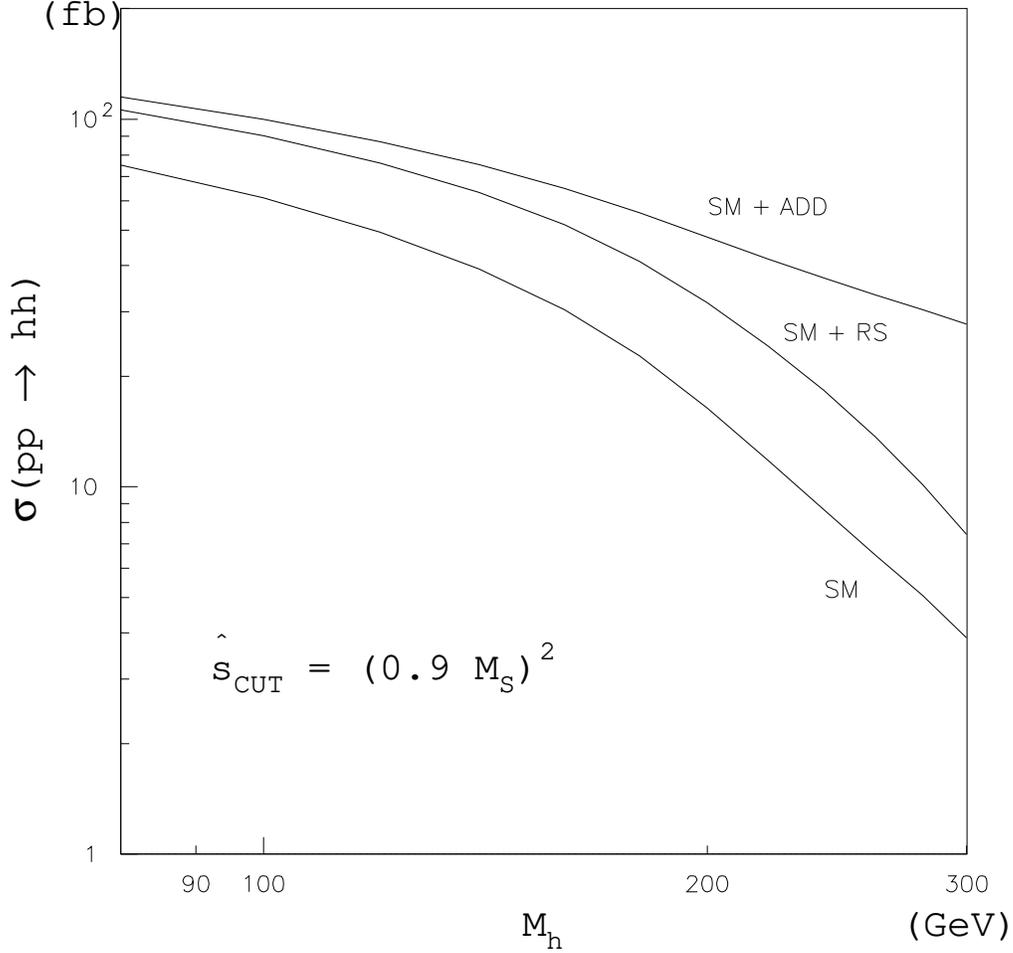,height=15cm}\hss} 
\caption{ The total cross section of the Higgs pair production  
as a function of the Higgs mass 
at the LHC with $\sqrt{s}=14$ TeV in the SM, the ADD and the RS cases. 
The string scales are set to be $M_S=2.5 $ TeV 
and $\Lm_\pi=3$ TeV for the ADD and the RS cases, respectively. 
The upper bounds in $M_{h h}<0.9\, M_S(\Lm_\pi)$ 
are employed. 
} 
\label{fig2} 
\end{figure} 
\end{center} 
 
\newpage 
 
\begin{center} 
\begin{figure}[htb] 
\hbox to\textwidth{\hss\epsfig{file=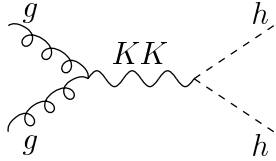}\hss} 
\caption{ The Feynman diagrams of the $g g \to h h $ 
                in the ADD and the RS models. 
} 
\label{fig3} 
\end{figure} 
\end{center}

\newpage 
 
\begin{center} 
\begin{figure}[htb] 
\hbox to\textwidth{\hss\epsfig{file=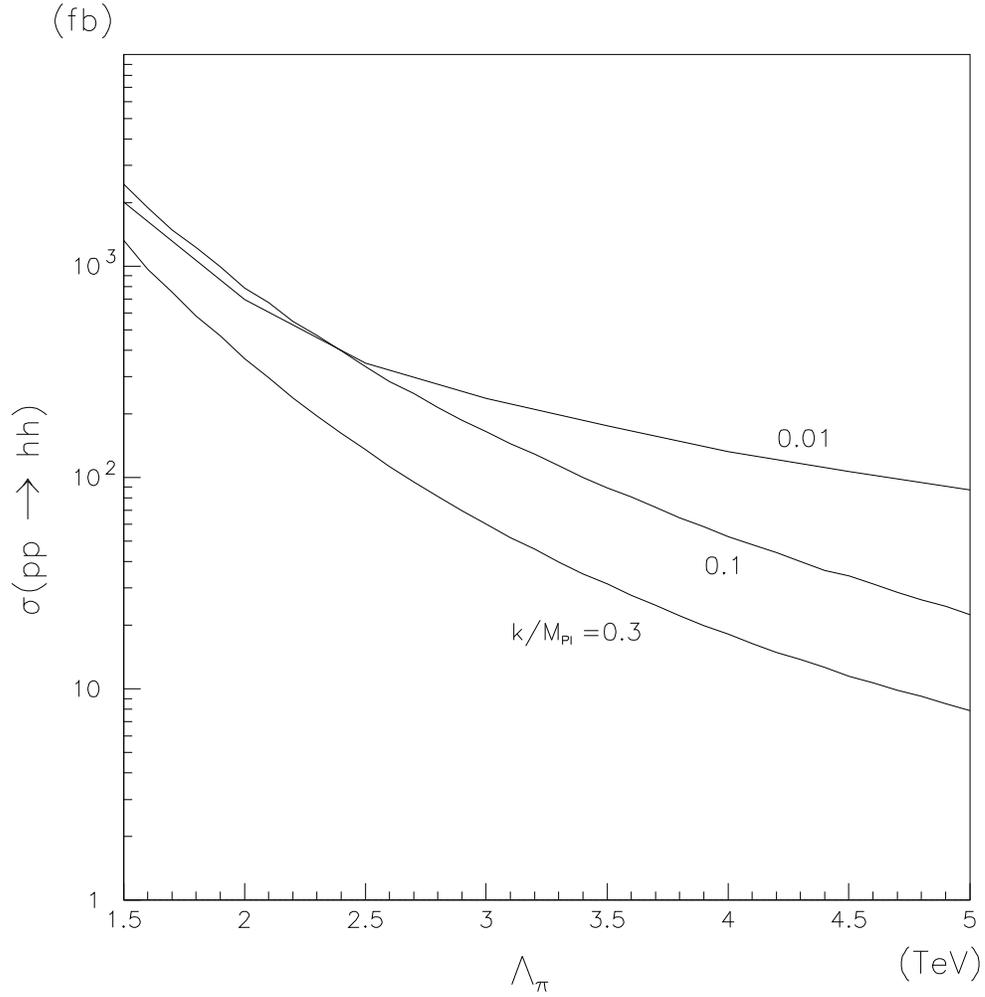,height=15cm}\hss} 
\caption{ The total cross sections only with the RS effects 
	as a function of $\Lambda_\pi$, 
        considering three values of the ratio 
        $ k/{M_{\rm Pl}}=0.01$, $0.1$ and $0.3$. 
} 
\label{fig4} 
\end{figure} 
\end{center} 
 
\newpage 
 
\begin{center} 
\begin{figure}[htb] 
\hbox to\textwidth{\hss\epsfig{file=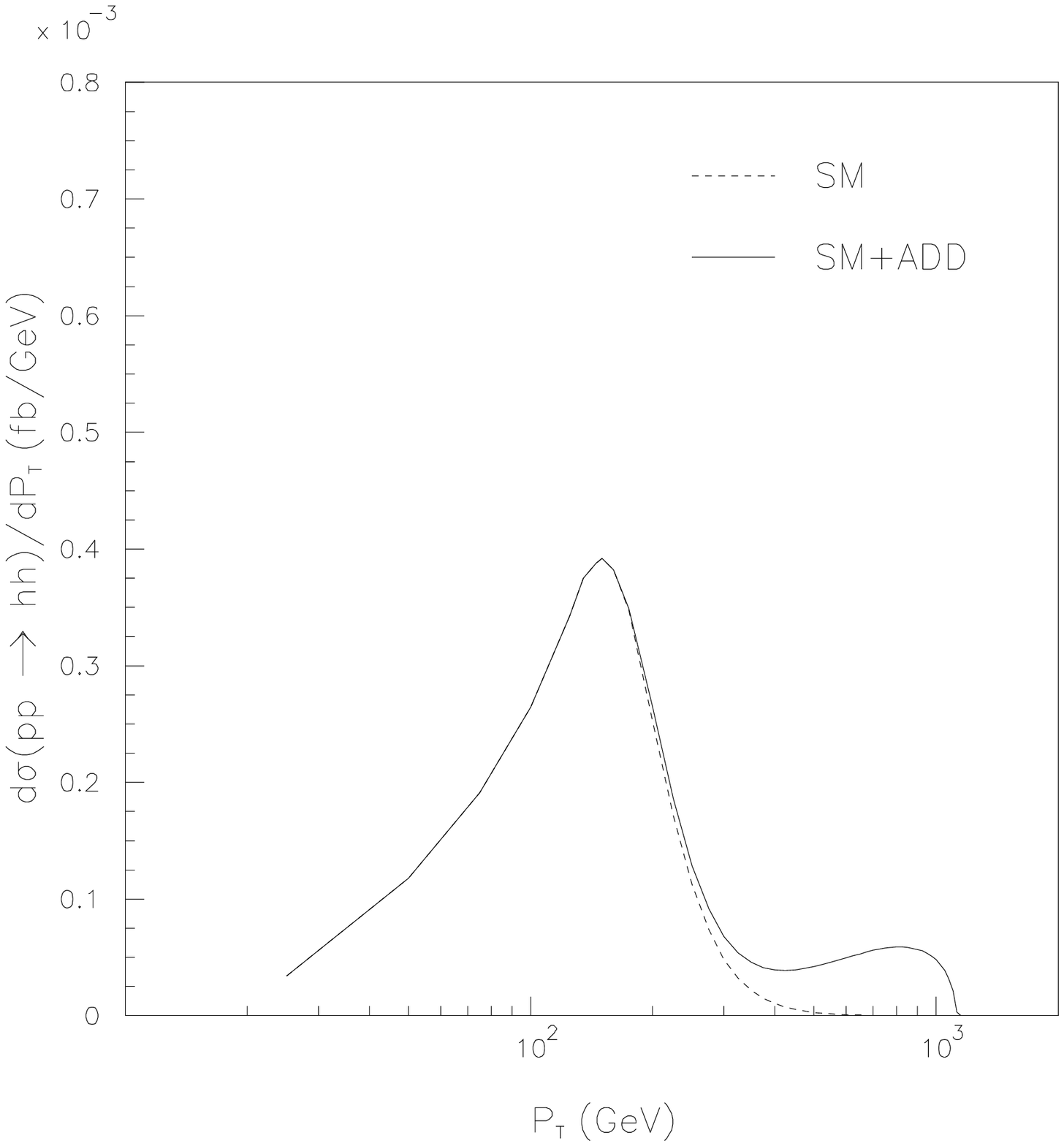,height=15cm}\hss} 
\caption{ The $p_{_T}$ distributions of the Higgs pair production 
        in the SM, and the ADD cases. 
} 
\label{fig5} 
\end{figure} 
\end{center} 
 
\newpage 
 
\begin{center} 
\begin{figure}[htb] 
\hbox to\textwidth{\hss\epsfig{file=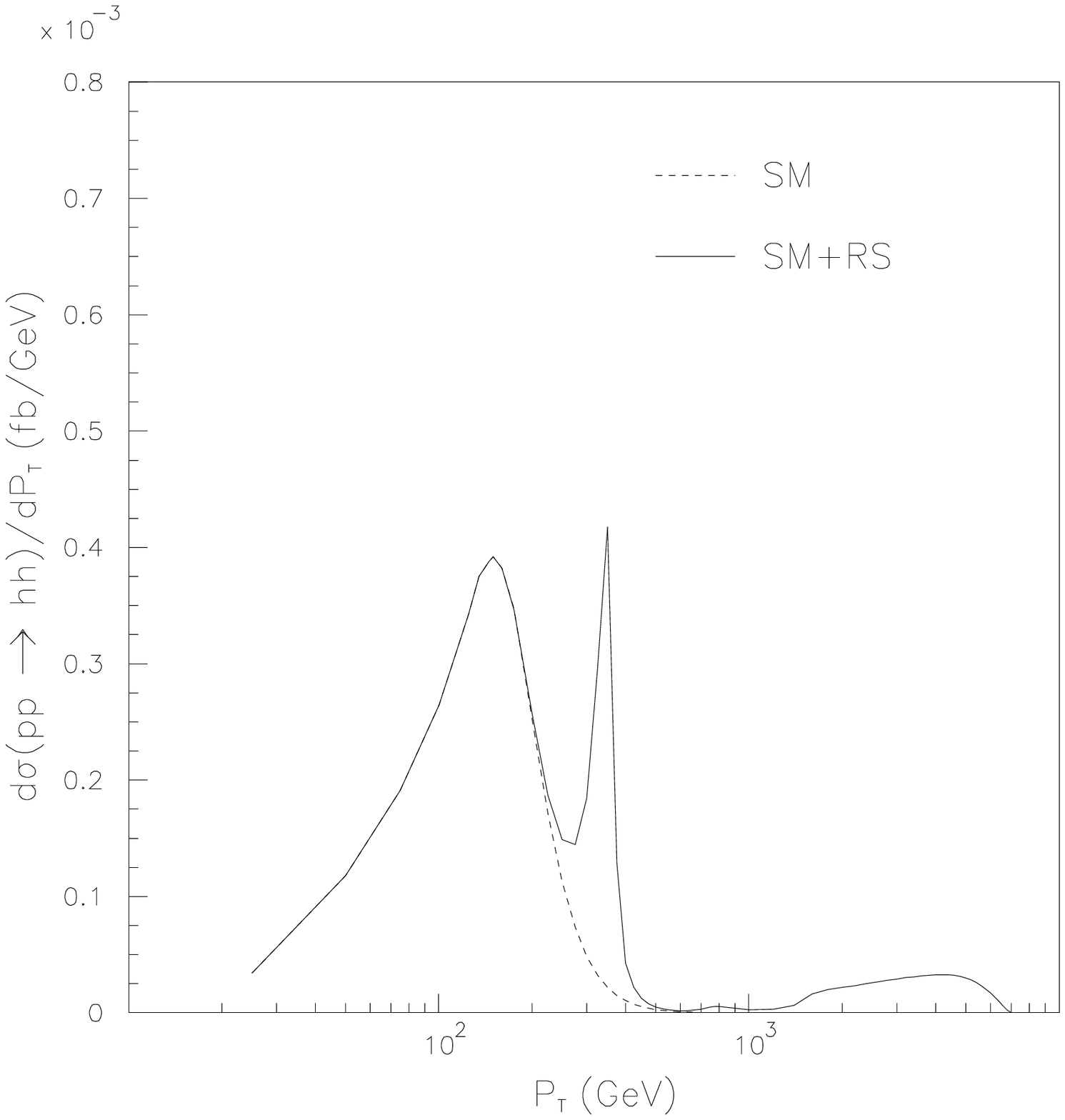,height=15cm}\hss} 
\caption{ The $p_{_T}$ distributions of the Higgs pair production 
        in the SM and the RS cases. 
} 
\label{fig6} 
\end{figure} 
\end{center} 
 
\newpage 
 
\begin{center} 
\begin{figure}[htb] 
\hbox to\textwidth{\hss\epsfig{file=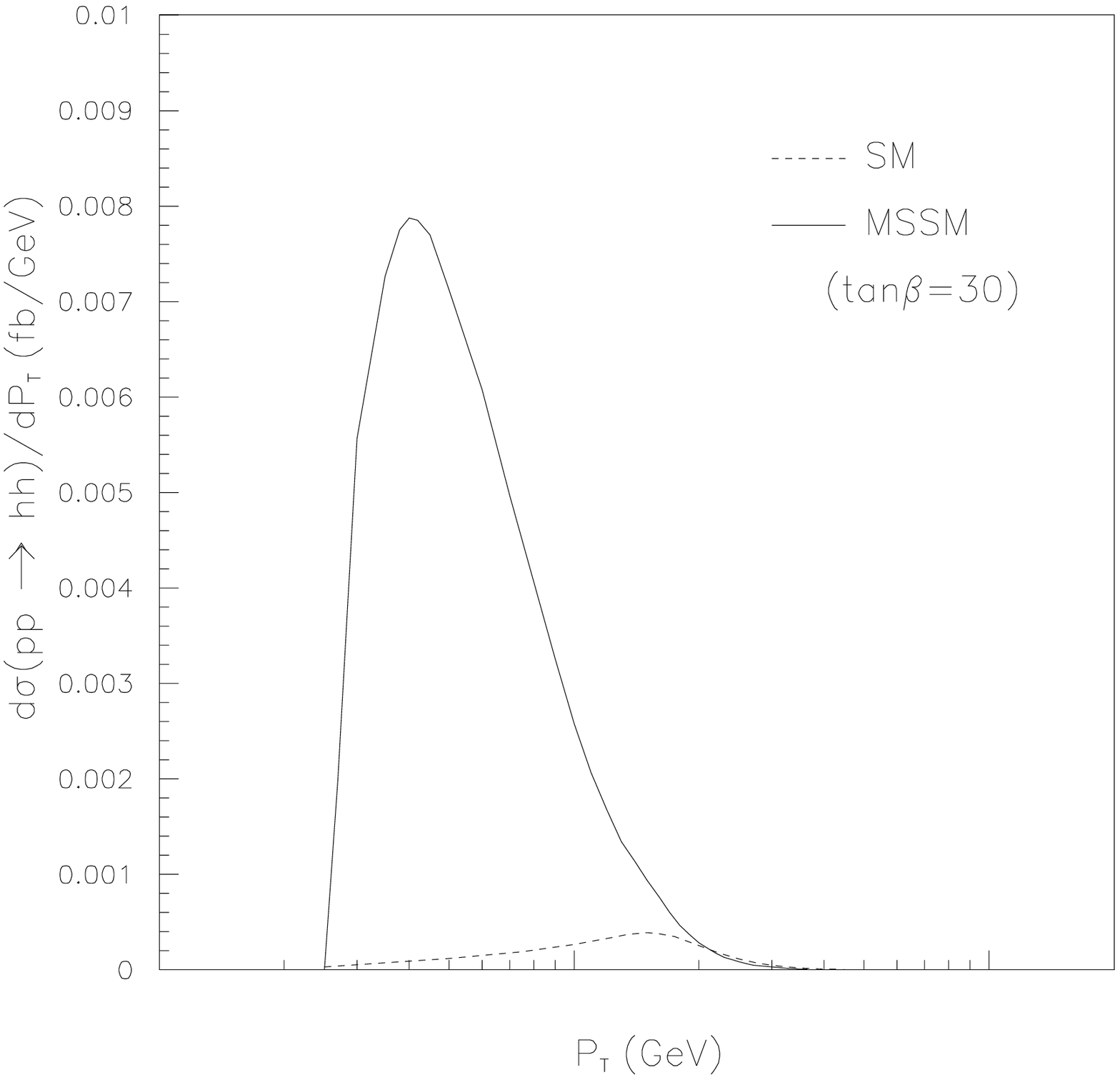,height=15cm}\hss} 
\caption{ The $p_{_T}$ distributions of the Higgs pair production 
        in the SM, and the MSSM. 
} 
\label{fig7} 
\end{figure} 
\end{center}

\newpage 
 
\begin{center} 
\begin{figure}[htb] 
\hbox to\textwidth{\hss\epsfig{file=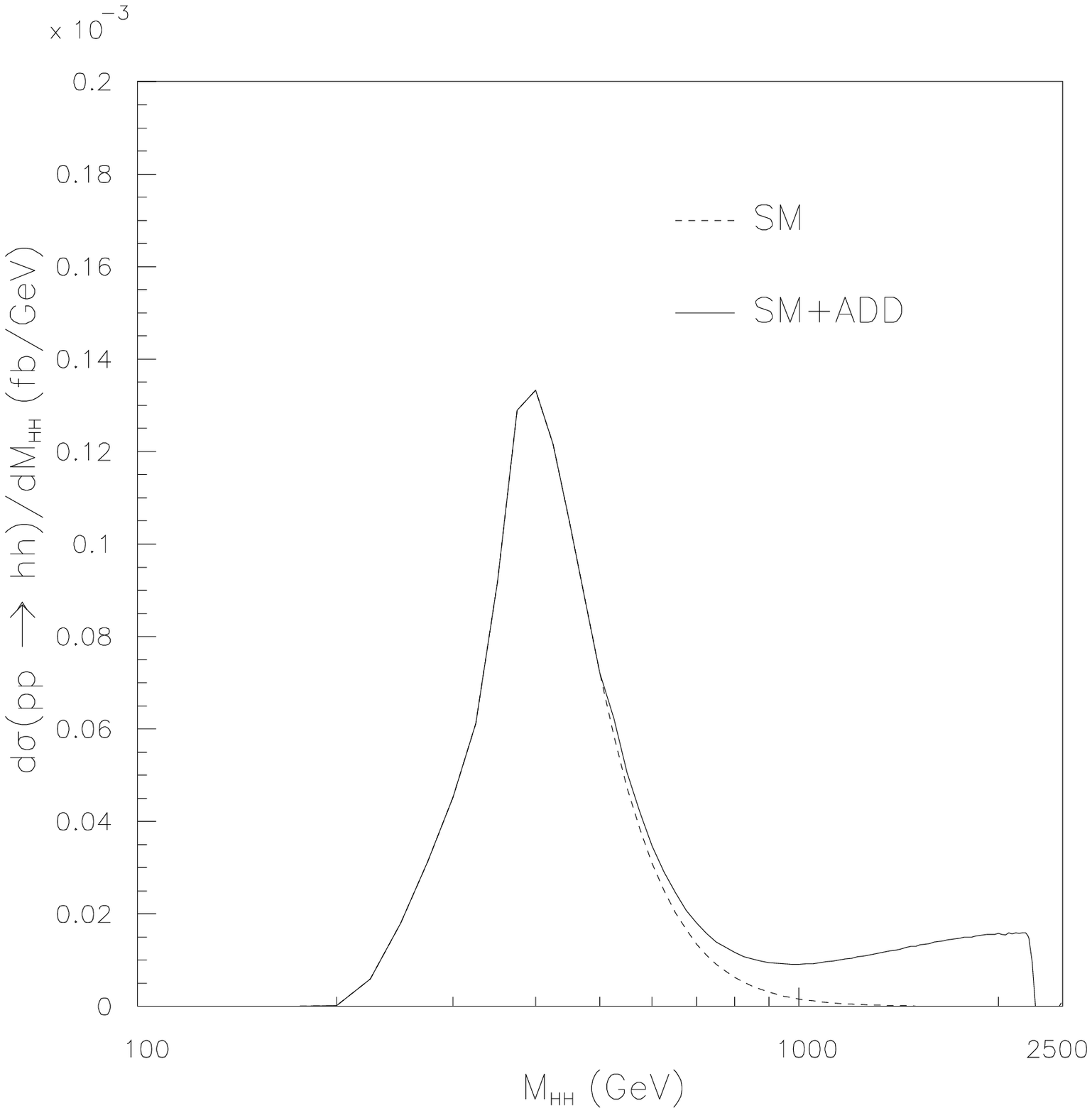,height=15cm}\hss} 
\caption{ The $M_{hh}$ distributions of the Higgs pair production 
        in the SM and the ADD cases. 
} 
\label{fig8} 
\end{figure} 
\end{center} 
 
\newpage 
 
\begin{center} 
\begin{figure}[htb] 
\hbox to\textwidth{\hss\epsfig{file=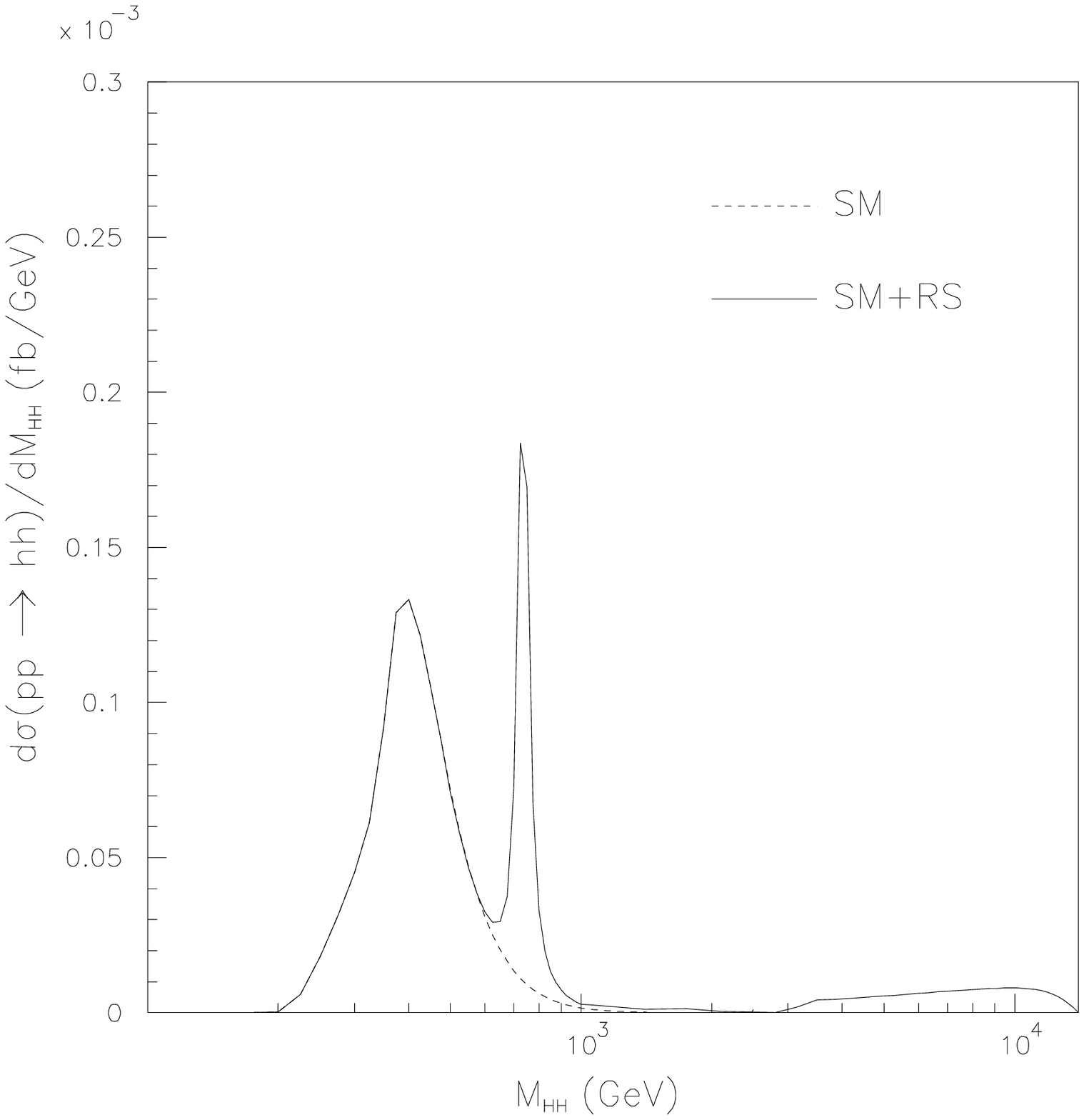,height=15cm}\hss} 
\caption{ The $M_{hh}$ distributions of the Higgs pair production 
        in the SM and the RS cases. 
} 
\label{fig9} 
\end{figure} 
\end{center} 
 
\newpage 
 
\begin{center} 
\begin{figure}[htb] 
\hbox to\textwidth{\hss\epsfig{file=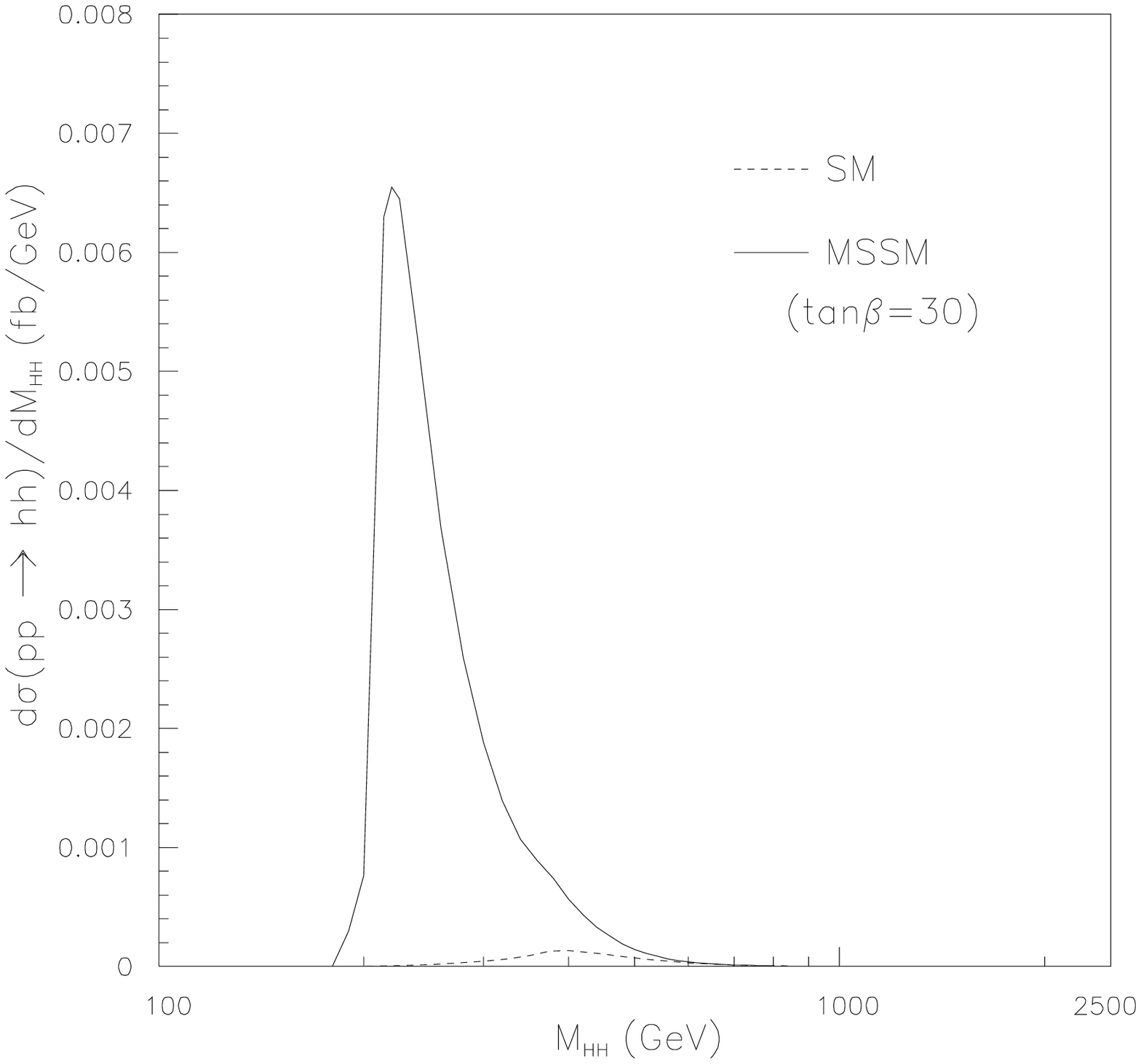,height=15cm}\hss} 
\caption{ The $M_{hh}$ distributions of the Higgs pair production 
        in the SM and the MSSM. 
} 
\label{fig10} 
\end{figure} 
\end{center} 
\newpage 
 
\begin{center} 
\begin{figure}[htb] 
\hbox to\textwidth{\hss\epsfig{file=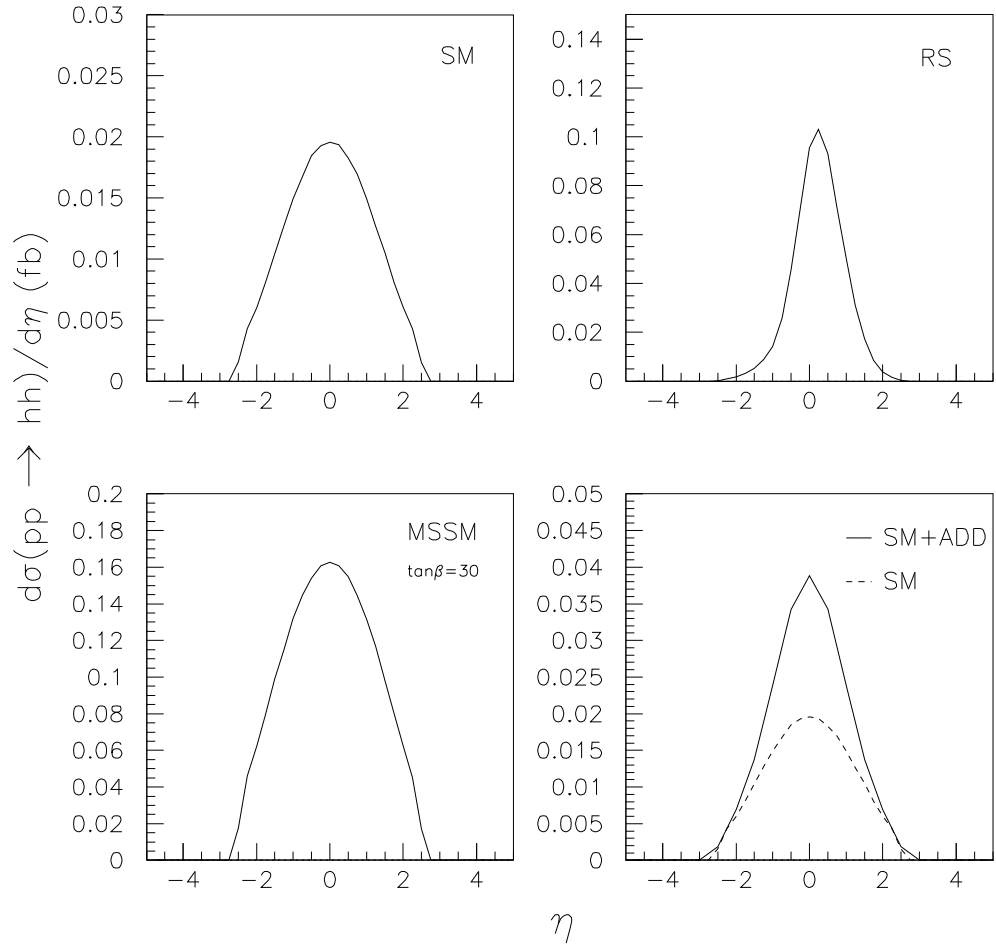,height=15cm}\hss} 
\caption{ The $\eta$ distributions  
                in the SM, the MSSM, the RS and the ADD cases. 
} 
\label{fig11} 
\end{figure} 
\end{center}

\end{document}